\documentclass[preprint2]{aastex}

\slugcomment{To be submitted to \emph{Astrophysical Journal}}

\begin{document}

\title{On the very high energy ($>25GeV$) pulsed emission in the Crab pulsar.}

\author{Machabeli George\altaffilmark{1} and Osmanov Zaza\altaffilmark{2}}
\affil{Georgian National Astrophysical Observatory, Chavchavadze
State University, Kazbegi 2a, 0106, Tbilisi, Georgia}
\altaffiltext{1}{g.machabeli@astro-ge.org}

\altaffiltext{2}{z.osmanov@astro-ge.org}

\begin{abstract}
We have examined the recently detected very high energy (VHE) pulsed
radiation from the Crab pulsar. According to the observational
evidence, the observed emission ($>25GeV$) peaks at the same phase
with the optical spectrum. Considering the cyclotron instability, we
show that the pitch angle becomes non-vanishing leading to the
efficient synchrotron mechanism near the light cylinder surface. The
corresponding spectral index of the emission equals $-1/2$. By
studying the inverse Compton scattering and the curvature radiation,
it is argued that the aforementioned mechanisms do not contribute to
the VHE radiation detected by MAGIC.
\end{abstract}


\section{Introduction}

For studying the high energy radiation of pulsars, two major
mechanisms are used: the synchrotron process \citep{pacini,shkl} and
the Inverse Compton Scattering (ICS) \citep{blan}. However, the
problem of identification of a location, where the radiation comes
from still is a matter of discussion. For solving this problem an
approach of the so-called polar cap model was proposed \citep{stur}.
According to this model, due to very strong electrostatic field,
particles are uprooted from the star's surface layer and accelerate
along the magnetic field lines \citep{rudsuth} inside a zone where
the electric field is nonzero (gap). This process leads to
relativistic energies of electrons, which in turn, create radiation.
Unfortunately, the particle energy inventory accumulated within the
gap is not enough to explain the observed high energy emission. For
solving this problem, to enlarge somehow the gap zone, series of
works have been invoked. \citet{arons} have considered an effect of
rectifying the curved magnetic field lines and they have shown that
this process leads to an increase of the gap size. \cite{hard} have
applied the method developed by \cite{arons} for studying the high
altitude emission from the pulsar slot gaps. A three-dimensional
model of optical and $\gamma$-ray emission from the slot gap
accelerator was examined and it has been predicted that the emission
below $200MeV$ must exhibit correlations in time and phase with the
radio band. Somewhat different mechanism leading to the effect of
the gap size increase was proposed by \cite{usov} where the
intermediate formation of positronium (electron-positron bound
state) was studied. Generally speaking, the space-time close to
neutron star's surface is slightly curved. Based on this fact,
\citet{musl} have considered an effect of creation od an additional
electric field due to the influence of the Kerr metric. All
aforementioned mechanisms cannot provide a significant increase of
the gap size to explain the observed high energy radiation from
pulsars. To solve this problem, the so-called outer gap model has
been proposed [see for example \citep{cheng1,cheng2}]. According to
this approach radiation is formed further out in the magnetosphere
of pulsars and mechanisms responsible for emission are: the
synchrotron, inverse Compton and curvature radiation processes
respectively.

Recently, MAGIC Cherenkov telescope has detected from the Crab
pulsar the VHE pulsed emission above $25GeV$ \citep{magic}. The most
peculiar feature of the observed radiation is coincidence of the
high energy and optical signals \citep{magic}. This indicates that
the polar cap models cannot be applied for explaining the MAGIC
observational data. In \citep{difus} we have examined the Crab
pulsar's recently observed VHE pulsed emission data. It is worth
noting that due to very small cooling timescales, particles rapidly
transit to the ground Landau state completely killing the subsequent
radiation. We have found that due to the cyclotron instability, the
optical spectrum is generated, which in turn via the quasi-linear
diffusion provokes the increase of the pitch angles, leading to the
synchrotron process with the spectral index $-1/2$. It has been
shown that the emission in the optical and high energy ($>25GeV$)
bands originate from the well localized regions, leading to the
observational fact that the signals peak with the same phases. As we
have already noted, the analysis of the observational data indicates
that the curvature radiation and the inverse Compton mechanism must
be excluded from the possible emission mechanisms. Generally
speaking, for very strong corotating magnetic fields, the particles
a) move along curved trajectories and b) accelerate up to
relativistic energies. Therefore, it is of great importance to
understand why the curvature radiation and the ICS are not involved
in the process of emission.

The paper is organized as follows. In Section 2 we consider the
synchrotron, Inverse Compton and curvature radiation mechanisms, in
Sect. 3 we present our results and in Sect. 4 we summarize them.

\section{Main consideration} \label{sec:consid}

In this section we are gong to consider three major emission
mechanisms: the synchrotron process, the ICS and the curvature
radiation respectively.

\subsection{Synchrotron emission} \label{sec:consid}

As we have already mentioned, timescales of transit of particles to
the ground Landau state is very small and electrons very soon start
moving along the magnetic field lines without emission.
\cite{machus1} have found that the situation changes due to the
cyclotron instability, leading to creation of nonzero pitch angles
and the corresponding radiation process. The method developed in
(\cite{machus1}) was applied by us in \citep{difus}. We have shown
that the quasi-linear duffusion excites the transverse and
longitudinal-transverse waves:
\begin{equation}\label{disp1}
\omega_t = kc\left(1-\delta_1\right),
\end{equation}
\begin{equation}\label{disp2}
\omega_{lt} = k_{_{\|}}c\left(1-\delta_1-\delta_2\right)
\end{equation}
where $k$ is the modulus of the wave vector, $k_{_{\|}}$ and
$k_{\perp}$ are the wave vector's longitudinal (parallel to the
magnetic field) and transverse (perpendicular to the magnetic field)
components respectively, $c$ is the speed of light and
\begin{equation}\label{delt}
\delta_{1} = \frac{\omega_p^2}{4\omega_B^2\gamma_p^3}, \;\;\;\;
\delta_{2} = \frac{k_{\perp}^2 c^2}{16\omega_p^2\gamma_p},
\end{equation}
were, $\omega_p \equiv \sqrt{4\pi n_pe^2/m}$ is the plasma
frequency, $\omega_B\equiv eB/mc$ is the cyclotron frequency, $e$
and $m$ are electron's charge and the rest mass respectively and
$n_p$ is the plasma density. According to a model developed by
\cite{difus}, we consider the plasma composed of two components: a)
the plasma component with the Lorentz factor, $\gamma_p$ and b) the
beam component with the Lorentz factor, $\gamma_b$.

\cite{kmm} have shown that the aforementioned modes generate if the
cyclotron resonance regime
\begin{equation}\label{cycl}
\omega - k_{_{\|}}V_{_{\|}}-k_xu_x\pm\frac{\omega_B}{\gamma_b} = 0,
\end{equation}
takes place. $u_x\equiv cV_{_{_{\|}}}\gamma_b/\rho\omega_B$ is the
drift velocity, $V_{_{\|}}$ is the longitudinal velocity component
and $\rho$ is the curvature radius of field lines.

For the parameters of the Crab pulsar, $P\approx 0.033sec$
$R_s\approx 10^6cm$, $n_{ps}\approx 1.4\times 10^{19}cm^{-3}$,
$B_s\approx 7\times 10^{12}G$, $\gamma_b\approx 10^8$, from Eqs.
(\ref{disp1}-\ref{cycl}) one can show that the development of the
cyclotron instability occurs in the optical band ($\sim 10^{15}$)
close to the light cylinder zone. $P$ denotes the pulsar's period,
$R_s$ - its radius, and $n_{ps}$ and $B_s$ are the plasma density
and the magnetic field induction respectively, close to the star.

When electrons emit in the synchrotron regime, they experience the
so-called radiative forces \citep{landau}:
$$F_{\perp} = -\alpha\psi(1 + \gamma_b^2\psi^2),$$
\begin{equation}\label{f}
F_{_{\|}} = -\alpha\gamma_b^2\psi^2,\;\;\;\;\;\alpha =
\frac{2}{3}\frac{e^2\omega_B^2}{c^2},
\end{equation}
where $\psi$ is the pitch angle. These forces try to decrease the
pitch angle contrary to the quasi-linear diffusion, which tends to
widen the value of $\psi$. The dynamical process saturates when
these two processes balance each other. As it has been shown
\citep{difus,malmach}, when $\gamma\psi\gg 1$ the kinetic equation
in the quasi-stationary regime ($\partial /\partial t=0$) reduces to
$$\frac{\partial\left[ F_{_\parallel}f\right]}{\partial
p_{_\parallel}} +
\frac{1}{p_{_\parallel}\psi}\frac{\partial\left[\psi
F_{\perp}f\right]}{\partial\psi}
=\frac{1}{\psi}\frac{\partial^2}{\partial\psi\partial
p_{_\parallel}}\left(D_{\perp_{\parallel}}\frac{\partial
f}{\partial\psi}\right)+$$

\begin{equation}\label{kinet}
+\frac{1}{\psi}\frac{\partial}{\partial\psi}\left[\psi\left
(D_{\perp\perp}\frac{\partial}{\partial\psi} +
D_{\perp_{\parallel}}\frac{\partial}{\partial
p_{_\parallel}}\right)f\right],
\end{equation}
where $f = f(\psi,p{_\parallel})$ is the distribution function of
particles, $p_{_\parallel}$ is the longitudinal momentum,
\begin{equation}\label{dif}
D_{\perp\perp}\approx -\frac{\pi^2 e^2n_bc}{2\omega},\;\;\;\;\;\;
D_{\perp_{\parallel}}\approx \frac{\pi^2
e^2n_b\omega_B}{2mc\gamma^2\omega^2},
\end{equation}
are the diffusion coefficients and $n_b = B/(Pce)$ is the density of
the beam component. This equation can be solved easily if one
expresses the distribution function as
$\chi(\psi)f(p_{_\parallel})$. Then, after substituting this
expression into Eq. (\ref{kinet}) one gets following
\citep{difus,nino}:
\begin{equation}\label{chi} \chi(\psi) = C_1e^{-A\psi^4},
f(p_{_\parallel}) = \frac{C_2}{\left(\alpha\bar{\psi}^2\gamma_b^2
-\frac{\pi^2e^2\bar{\psi}n_bc}{\gamma_b}\right)},
\end{equation}
where
\begin{equation}\label{A}
A\equiv \frac{4e^6B^4P^3\gamma_p^4}{3\pi^3m^5c^7\gamma_b},
\end{equation}
and
\begin{equation}\label{pitch}
\bar{\psi}
 = \frac{\int_{0}^{\infty}\psi\chi(\psi)d\psi}{\int_{0}^{\infty}\chi(\psi)d\psi}
\approx \frac{0.5}{\sqrt[4]{A}}.
\end{equation}
is the mean value of the pitch angle.

Relativistic electrons moving in the magnetic field emit photons
with energies  expressed by $\epsilon\approx 1.2\times
10^{-17}B\gamma^2\sin\psi (GeV) $\citep{Lightman}, which after
applying Eq. (\ref{pitch}) reduces to:
\begin{equation}
\label{eps1} \epsilon_{syn}(GeV)\approx 6\times
10^{-18}\left(\frac{3\pi^3m^5c^7\gamma_b^9}{4P^3e^6\gamma_p^4
}\right)^{\frac{1}{4}}.
\end{equation}

According to the results of the MAGIC Cherenkov telescope
\citep{magic}, the VHE ($>25GeV$) pulsed radiation from the Crab
pulsar peaks with the same phase as the optical signal. This
circumstance is in a good agreement with our model, since, as we
have shown, the pitch angles increase due to the cyclotron
instability, which in its turn occurs in the optical frequency
ranges.

Let us apply Eqs. (\ref{pitch},\ref{eps1}) to the Crab pulsar. Then
by taking into account the following parameters, $R_s\approx
10^6cm$, $B_s\approx 7\times 10^{12}G$ and $\gamma_p \approx 3$, one
can show that the pitch angle is of the order of $10^{-5}$, which
guarantees the observed high energy emission if $\gamma_b\approx
3.2\times 10^8$. For the mentioned set of parameters one has
$\alpha\bar{\psi}^2\gamma_b^2\gg\pi^2e^2\bar{\psi}n_bc/\gamma_b$,
which reduces the distribution function to
$f(p_{_\parallel})\propto\gamma_b^{-2}$. On the other hand, the
spectrum of the synchrotron radiation behaves as
$I_{\nu}\propto\nu^{-\frac{\beta-1}{2}}$ \citep{ginz}, where $\beta$
describes the particle distribution function,
$f\propto\gamma^{-\beta}$. This indicates that, for the case
considered in the present paper ($\beta = 2$) the spectral index of
the VHE synchrotron emission equals $-1/2$.

Therefore, as we see, the synchrotron emission can explain the
observed VHE radiation, and as we have already seen, for this
purpose the particles must have very high Lorentz factors. On the
other hand, these particles will inevitably encounter soft photons,
which in turn can also create the high energy radiation via the ICS.
But in this case the emission will not be localized contrary to the
observational evidence, indicating that for some reason the ICS is
not involved in the process of the detected emission. The next
subsection is dedicated to this particular problem.

\subsection{Compton scattering}

It is well known that when a photon with energy $\epsilon$
encounters a relativistic electron, under certain conditions photons
might gain energy. The corresponding frequency after scattering is
given by \citep{Lightman}:
\begin{equation}\label{ics1}
\omega' = \omega\frac{1-\beta\cos\theta}{1-\beta\cos\theta'+
\frac{\hbar\omega}{\gamma mc^2}\left(1-\cos\theta''\right)},
\end{equation}
where $\omega$ is the frequency before scattering, $\beta\equiv
\upsilon /c$, $\theta=(\widehat{{\bf PK}})$, $\theta'=(\widehat{{\bf
PK'}})$, $\theta''=(\widehat{{\bf KK'}})$. By ${\bf K}$ and ${\bf
K'}$ we denote the three momentum of the photon before and after
scattering respectively. The momentum of relativistic electrons
before scattering is denoted by ${\bf P}$.

Since, according to the observational evidence, we observe the well
localized pulses of high energy emission, therefore the angle,
$\theta$ must be very small. On the other hand, analyzing the
excitation of oblique waves in a relativistic electron-positron
plasma one can argue that the pitch angle has to be extremely low
\citep{vol}. Then, Eq. (\ref{ics1}) reduces to
\begin{equation}\label{ics2}
\omega' \approx\omega\frac{1-\beta}{1-\beta\cos\theta'+
\frac{\hbar\omega}{\gamma mc^2}\left(1-\cos\theta''\right)}.
\end{equation}
We will study two principally different cases: (a) $\cos\theta'\ll
1$ and (b) $\cos\theta'\sim 1$. In the first case, we have
\begin{equation}\label{ics3}
\omega' \approx\frac{\omega}{2\gamma^2}\times\frac{1}{1+
\frac{\hbar\omega}{\gamma mc^2}\left(1-\cos\theta''\right)},
\end{equation}
where we have taken into account the following approximate relation,
$1-\beta \approx 1/2\gamma^2$ for $\beta\sim 1$. From this
expression we see that for all physical quantities the frequency
after the scattering is less than that of before the scattering and
therefore, there is no possibility of increasing $\omega$ to the VHE
band.

By considering the second limit, Eq.  (\ref{ics2}) reduces to
\begin{equation}\label{ics4}
\omega' \approx \omega\frac{1}{1+
\frac{2\hbar\omega\gamma}{mc^2}\left(1-\cos\theta''\right)},
\end{equation}
which, as in the previous case, leads to the similar result,
$\omega'<\omega$.

This investigation shows that the ICS cannot provide the VHE
radiation from the Crab pulsar detected by MAGIC \citep{magic}.

\subsection{Curvature radiation}

Since particles are moving along the curved magnetic field lines
continuously, they will emit the curvature radiation. On the other
hand, in Eq. (\ref{kinet}) we have neglected a term corresponding to
the curvature emission. This in its turn, means that the following
ratio
\begin{equation}
\label{rat} \eta\equiv \frac{\epsilon_{cur}}{\epsilon_{syn}},
\end{equation}
where \citep{rudsuth}
\begin{equation}
\label{cur} \epsilon_{cur} =
\frac{3\hbar}{2}\gamma_b^3\times\frac{c}{\rho},
\end{equation}
and $\rho$ is the curvature radius of magnetic field lines, must be
less than one. By applying Eqs. (\ref{eps1},\ref{rat},\ref{cur}),
one can show that for typical  magnetospheric parameters of the Crab
pulsar close to the light cylinder, the aforementioned ratio is
negligible only if the curvature radius exceeds the light cylinder
radius, $R_{lc}$, approximately by three orders of magnitude.

Generally speaking, since no contribution in emission comes from the
closed magnetic field lines, we consider the open ones. On the other
hand, in the dipolar field the region of almost straight field lines
is just a tiny fraction of the emission area, leading to a
negligible value of the high energy luminosity.


If the beam component particles move along curved field lines, they
experience the so-called curvature drift with the velocity:
\begin{equation}
\label{u} u_{b}= \frac{\gamma_{b_0} v_{_\parallel}^2}{\omega_{B_{b}}
\rho},
\end{equation}
where $\omega_{B_b} = eB_0/mc$; and $B_0$ is the background magnetic
field induction. This velocity will eventually create the drift
current, $J_{dr} = en_bu_b$ which in turn, via the Maxwell equation
\begin{equation}
\label{ind1} \left({\bf \nabla\times B}\right)_x =
\frac{4\pi}{c}{J_{_dr}},
\end{equation}
can create the toroidal magnetic field (by $x$ we denote a direction
of the drift current. See Fig. \ref{fig}). This current is evidently
less than the Goldreich-Julian (GJ) current $J_{GJ} = en_b c$, since
$u_b\ll c$. But on the other hand, the GJ current creates the
corresponding magnetic field, $B_r\approx 4\pi J_{GJ}R_n/c$, where
$R_n\approx B(R)/B'(R) = R/3$ ($B'\equiv dB/dR$) is the length scale
of the spatial inhomogeneity of the magnetic field. If we assume a
dipolar configuration, then, by taking the value of the GJ density,
$n_b\approx \Omega B/(2\pi ec)$, into account, one can show that the
toroidal magnetic field equals $\frac{2R}{3R_c}B$. Inside the light
cylinder ($R< R_c$), this value is less than the background magnetic
field-$B$ and therefore such a toroidal magnetic field will be
unable to rectify the twisted magnetic field lines. This implies
that the curvature drift current, which is less than that of the GJ,
cannot contribute to the process of rectifying the field lines.
However, in spite of that the drift current is not the source of the
toroidal component, $B_r$, it is a trigger mechanism for generatign
the perturbed current, $J_1= e(n_b^0v^1_{b_x}+n_b^1u_b)$ responsible
for the creation of $B_r$ (see Eq. (\ref{indp})), where by upper
script $"1"$ we denote the perturbed quantities. The source of the
instability of current and the resulting magnetic field is the
pulsar's rotational energy and the process is achieved via the
parametrically excited curvature drift waves. The corresponding
increment of the curvature drift instability can be presented by
(see Appendix, for more details see \citep{mnras,forcefree}):
\begin{equation}
\label{gama} \Gamma \approx
\left(-\frac{3}{2}\frac{\omega^2_{b}}{\gamma_{b_0}}\frac{k_xu_{b}}{k_{\theta}c}\right)
^{1/2}\left|J_0\left(\frac{k_xu_{b}}
{4\Omega}\right)J_{0}\left(\frac{k_{\theta}c}{\Omega}\right)\right|,
\end{equation}
where $\omega_b$ is the beam component plasma frequency and
$\gamma_{b0}$ - the Lorentz factor in an unperturbed state. $k_x$
and $k_{\theta}$ are the wave vector's components and $\Omega$ is
the angular velocity of rotation.

By considering the typical magnetospheric parameters for the Crab
pulsar close to the light cylinder, $P\approx 0.033s$, $\gamma_{b0}
= 10^8$ and examining the perturbation lengthscale $\lambda\approx
10^8cm$ \citep{forcefree}, one can see that the increment is of the
order of $1s^{-1}$. Comparing this value with the Crab pulsar's
slowdown rate, $4.2\times 10^{-13}s^{-1}$, we see that the
instability growth rate exceeds by many orders of magnitude the
slowdown rate, indicating that the mentioned instability is
extremely efficient.

It is worth noting that we have three types of the open field lines:
(a) curved field lines which pass ahead of the rotation; (b) a tiny
fraction of almost straight field lines and (c) curved field lines,
lagging behind the rotation.

If the initial perturbation of the toroidal magnetic field satisfies
the condition $B_r>0$, then such a perturbation will rectify all
field lines which initially pass ahead of the rotation (suppose the
clockwise rotation of the system) and will twist even more the
magnetic field lines, which initially lag behind the rotation. In
the case, $B_r<0$, the situation is opposite: the field lines
initially lagging behind the rotation will be rectified. At this
stage the curvature becomes infinity and as we see from Eq.
(\ref{u}), the drift velocity tends to zero, saturating (killing)
the instability.

The investigation shows that, the curvature drift instability
provides necessary conditions for an efficient mechanism of
rectifying the field lines, leading to the negligible role of the
curvature radiation, confirming our assumptions leading to Eq.
(\ref{kinet}).

\section{Summary}\label{sec:summary}

\begin{enumerate}

      \item We have considered several emission mechanisms
      for explaining the recently detected VHE emission from the
      Crab pulsar.

      \item Studying the synchrotron mechanism, we have shown that
      due to the cyclotron instability
      efficiently developing on light cylinder scales,
      non-vanishing pitch angles are created, that
      leads to the efficient high energy synchrotron emission with
      the spectral index $-1/2$.

      \item The observational fact of the coincidence of signals in
      optical and high energy ($>25GeV$) intervals is in a good agreement with our
      model, in the framework of which, the cyclotron instability is
      excited in the optical spectra, which, via the synchrotron process
      leads to the high energy emission.

      \item Analyzing the inverse Compton scattering, we have found that
      for Crab pulsar's magnetospheric parameters
      even very energetic electrons are unable to produce the observed photon
      energies.

      \item Considering the curvature radiation, we show that due to the
      curvature drift instability, the magnetic field lines are rectified very
      efficiently. This in turn, leads to a negligible role of the
      curvature emission process in the observed VHE emission along the
      aforementioned rectified field lines.

      \end{enumerate}

As we see, the detected coincidence of VHE and optical signals is an
indirect confirmation of the fact that (a) both spectra is produced
by one source and (b) the only mechanism providing the detected high
energy radiation is the synchrotron mechanism. This means that we
observe the Crab pulsar towards these straight field lines, that is
the reason why we do not see the curvature radiation coming from the
twisted magnetic field lines.

\section*{Acknowledgments}
The research was supported by the Georgian National Science
Foundation grant GNSF/ST06/4-096.

\appendix
\section{Curvature drift instability}

In this section we study the process of rectifying the magnetic
field lines due to the parametrically excited curvature drift
instability. This instability is called parametric, because an
external force - centrifugal force, plays a role of a parameter,
changes in time and creates the instability. Generally speaking, the
presence of an external varying parameter generates the plasma
instability. The mechanism of energy pumping process from the
external alternating electric field into the electron-ion plasma is
quite well investigated in \citep{silin1,gal,max}. Instead of
considering the altering electric field, one can examine the
centrifugal force as a varying parameter \citep{incr1}.

\begin{figure}
  \resizebox{\hsize}{!}{\includegraphics[angle=0]{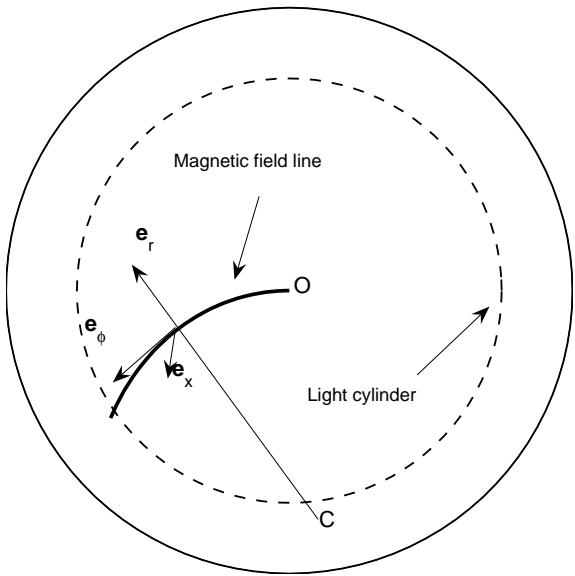}}
  \caption{Here we show geometry in which we consider our system of equations.
 By ${\bf e_{\phi}}$, ${\bf e_{r}}$ and ${\bf e_{x}}$ unit vectors are denoted, note that
 ${\bf e_{x}}\perp {\bf e_{r,\phi}}$. $O$ is the center of rotation and $C$ - the curvature center.}\label{fig}
\end{figure}

We start our consideration by supposing that the magnetic field
lines are almost straight with very small nonzero curvature (see
Fig. \ref{fig}). In this context we examine the field lines that are
open, and thus have the curvature radius exceeding the light
cylinder one, maximum by one order. Therefore, dynamics of
particles, governing the overall picture of the curvature drift
instability, can be studied, assuming that field lines are almost
straight. In the framework of the paper we suppose that the plasma
flow consists of two components: the plasma component composed of
electrons and positrons ($e^{\pm}$); and, the so-called, beam
component ($b$) composed of relativistic electrons. It is well known
that the dynamics of plasma particles moving along the straight
co-rotating magnetic field lines is described by the Euler equation:
\citep{incr1}:
\begin{equation}
\label{eul} \frac{\partial{\bf p_{\alpha}}}{\partial t}+({\bf
v_{\alpha}\nabla)p_{\alpha}}=
-c^2\gamma_{\alpha}\xi{\bf\nabla}\xi+\frac{e_{\alpha}}{m}\left({\bf
E}+ \frac{1}{c}\bf v_{\alpha}\times\bf B\right),
\end{equation}
where $\xi\equiv \sqrt{1-\Omega^2R^2/c^2}$, $R$ is the coordinate
along the straight field lines; ${\bf p_{\alpha}}$, ${\bf
v_{\alpha}}$, and $e_{\alpha}$ are the momentum (normalized to the
particle's mass), the velocity and the charge of
electrons/positrons, respectively; $\alpha=\{e^{\pm},b\}$ denotes
the sort of particles and ${\bf E}$ and ${\bf B}$ are the electric
and the magnetic field induction respectively. The continuity
equation:
\begin{equation}
\label{cont} \frac{\partial n_{\alpha}}{\partial t}+{\bf
\nabla}(n_{\alpha}{\bf v_{\alpha}})=0,
\end{equation}
and the induction equation:
\begin{equation}
\label{ind} {\bf \nabla\times B} = \frac{1}{c}\frac{\partial {\bf
E}}{\partial t}+\frac{4\pi}{c}\sum_{\alpha=e^{\pm},b}{\bf
J_{\alpha}},
\end{equation}
(where $n_{\alpha}$ and ${\bf J_{\alpha}}$ are the density and the
current, respectively) complete the set of equations for $n,{\bf
v},{\bf E}$ and ${\bf B}$.

In the leading state the plasma is in the frozen-in condition: ${\bf
E}_0+ \frac{1}{c}{\bf v}_{0\alpha}\times{\bf B}_0=0$, then, one can
show that the solution to the Euler equation in Eq. (\ref{eul}) for
ultra relativistic particle velocities in the leading state is given
by \citep{mr}:
\begin{equation}
\label{vr} v^{0}_{\theta}\equiv v_{_\parallel} = c\,\cos(\Omega t +
\varphi),
\end{equation}
where $v_{_\parallel}$ is the velocity component along the magnetic
field lines and $\varphi$ is the initial phase of each particle.

For solving the set of Eqs. (\ref{eul}-\ref{ind}), we will linearize
it assuming that, in the zeroth order of approximation, the flow has
the longitudinal velocity satisfying Eq. (\ref{vr}) and also drifts
along the $x$-axis driven by the curvature of magnetic field lines
(see Fig. \ref{fig}):
\begin{equation}
\label{drift} u_{\alpha}= \frac{\gamma_{\alpha_0}
v_{_\parallel}^2}{\omega_{B_{\alpha}} \rho},
\end{equation}
where $u_{\alpha}$ is the drift velocity; $\omega_{B_{\alpha}} =
e_{\alpha}B_0/mc$; and $B_0$ is the background magnetic induction.

Let us expand the physical quantities up to the first order terms:
\begin{equation}
\label{expansion} \Psi\approx \Psi^0 + \Psi^1,
\end{equation}
where $\Psi\equiv \{n,{\bf v},{\bf p},{\bf E},{\bf B}\}$. Then if we
examine only the $x$ components of Eqs. (\ref{eul},\ref{ind}), and
express the perturbed quantities as follows:
\begin{equation}
\label{pert} \Psi^1(t,{\bf r})\propto\Psi^1(t) \exp\left[i\left({\bf
kr} \right)\right] \,,
\end{equation}
by taking into account that $k_{\theta}\ll k_x$ and $k_r = 0$, and
bearing in mind that $v^1_{r}\approx cE^1_x/B_{0}$, one can show
that Eqs. (\ref{eul}-\ref{ind}) reduce to the form:
\begin{equation}
\label{eulp} \frac{\partial p^1_{{\alpha}_x}}{\partial
t}-i(k_xu_{\alpha}+k_{\theta}v_{_\parallel})p^1_{{\alpha}_x}=
\frac{e_{\alpha}}{mc}v_{_\parallel}B^1_{r},
\end{equation}
\begin{equation}
\label{contp} \frac{\partial n^1_{\alpha}}{\partial
t}-i(k_xu_{\alpha}+k_{\theta}v_{_\parallel})n^1_{\alpha}=
ik_xn_{\alpha}^0v^1_{\alpha_x},
\end{equation}
\begin{equation}
\label{indp} -ik_{\theta}cB^1_{r} = 4\pi
\sum_{\alpha=e^{\pm},b}e_{\alpha}(n_{\alpha}^0v^1_{\alpha_x}+n_{\alpha}^1u_{\alpha}).
\end{equation}
According to the standard method \citep{mnras}, after expressing
$v^1_{\alpha_x}$ and $n^1_{\alpha}$ in the following way:
\begin{equation}
\label{anzp} v^1_{\alpha_x}\equiv V_{\alpha_x}e^{i{\bf
kA_{\alpha}(t)}},
\end{equation}
\begin{equation}
\label{anzn} n^1_{\alpha}\equiv N_{\alpha}e^{i{\bf kA_{\alpha}}(t)},
\end{equation}
\begin{equation}
\label{Ax} A_{\alpha_x}(t) = \frac{u_{\alpha}}{2\Omega}\left(\Omega
t + \varphi\right) + \frac{u_{\alpha}}{4\Omega}\sin[2\left(\Omega t
+ \varphi\right)],
\end{equation}
\begin{equation}
\label{Af} A_{\alpha_\theta}(t) = \frac{c}{\Omega}\sin(\Omega t+
\varphi),
\end{equation}
and substituting them into Eqs. (\ref{eulp},\ref{contp}), one can
get the expressions:
\begin{equation}
\label{vx} v^1_{{\alpha}_x} =
\frac{e_{\alpha}}{mc\gamma_{\alpha_0}}{\rm e}^{i{\bf
kA_{\alpha}}(t)}\int^t{\rm e}^{-i{\bf
kA_{\alpha}}(t')}v_{_\parallel}(t')B_{r}(t')dt',
\end{equation}
$$n^1_{\alpha} =
\frac{ie_{\alpha}n_{{\alpha}}^0k_x}{mc\gamma_{\alpha_0}}{\rm
e}^{i{\bf kA_{\alpha}}(t)}\int^tdt'\int^{t''}{\rm e}^{-i{\bf
kA_{\alpha}}(t'')}v_{_\parallel}(t'')B_{r}(t'')dt'',$$
\begin{equation}
\label{n}
\end{equation}
which combined with Eq. (\ref{indp}), lead to the following form:
$$ -ik_{\theta}cB^1_{r}(t)
=\sum_{\alpha=e^{\pm},b}\frac{\omega^2_{\alpha}}{\gamma_{\alpha_0}c}{\rm
e}^{i{\bf kA_{\alpha}}(t)}\int^t{\rm e}^{-i{\bf
kA_{\alpha}}(t')}v_{_\parallel}(t')B_{r}(t')dt'+ $$
\begin{equation}
\label{ind1}i\sum_{\alpha=e^{\pm},b}\frac{\omega^2_{\alpha}}{\gamma_{\alpha_0}c}k_xu_{\alpha}{\rm
e}^{i{\bf kA_{\alpha}}(t)}\int^tdt'\int^{t''}{\rm e}^{-i{\bf
kA_{\alpha}}(t'')}v_{_\parallel}(t'')B_{r}(t'')dt'',
\end{equation}
where $\omega_{\alpha} = e\sqrt{4\pi n_{\alpha}^0/m}$ represents the
plasma frequency. If we apply the following identity:
\begin{equation}
\label{bess} {\rm e}^{\pm ix\sin y}=\sum_s J_s(x){\rm e}^{\pm isy},
\end{equation}
to Eq. (\ref{ind1}), the latter will simplify to the following form:
$$B_{r}(\omega) =
-\sum_{\alpha=e^{\pm},b}\frac{\omega^2_{\alpha}}{2\gamma_{\alpha_0}k_{\theta}c}\sum_{\sigma
= \pm
1}\sum_{s,n,l,p}\frac{J_s(g_{\alpha})J_n(h)J_l(g_{\alpha})J_p(h)}{\omega
+ \frac{k_xu_{\alpha}}{2}+\Omega (2s+n) } \times$$ $$\times
B_{r}\left(\omega+\Omega
\left(2[s-l]+n-p+\sigma\right)\right)\left[1-\frac{k_xu_{\alpha}}{\omega
+ \frac{k_xu_{\alpha}}{2}+\Omega (2s+n)}\right]\times$$
$$\times {\rm e}^{i\varphi\left(2[s-l]+n-p+\sigma\right)}+$$
$$+\sum_{\alpha=e^{\pm},b}\frac{\omega^2_{\alpha}k_xu_{\alpha}}{4\gamma_{\alpha_0}k_{\theta}c}\sum_{\sigma,\mu
= \pm
1}\sum_{s,n,l,p}\frac{J_s(g_{\alpha})J_n(h)J_l(g_{\alpha})J_p(h)}{\left(\omega
+ \frac{k_xu_{\alpha}}{2}+\Omega (2[s+\mu]+n)\right)^2 } \times$$
\begin{equation}
\label{disp11} \times B_{r}\left(\omega+\Omega
\left(2[s-l+\mu]+n-p+\sigma\right)\right)\times {\rm
e}^{i\varphi\left(2[s-l+\mu]+n-p+\sigma\right)},
\end{equation}
where
$$g_{\alpha} = \frac{k_xu_{\alpha}}{4\Omega}, \;\;\;\;\;\;\;\;\;\;\;\;h =
\frac{k_{\theta}c}{\Omega}$$
and $J_s(x)$ ($s=0;\pm 1;\pm 2 \ldots$) is the Bessel function of
integer order \citep{abrsteg}.

In order to solve Eq. (\ref{disp11}), one has to examine similar
equations, rewriting Eq. (\ref{disp11}) (with shifted arguments) for
$B_{r}(\omega\pm\Omega)$, $B_{r}(\omega\pm 2\Omega)$, etc.. This
implies that we have to solve the system with the infinite number of
equations, making the problem impossible to handle. Therefore, the
only way is to consider the physics close to the resonance
condition, that provides the cutoff of the infinite row in Eq.
(\ref{disp11}) ad makes the problem solvable \citep{silin}.

Studying the resonance condition of Eq. (\ref{disp11}), one can
derive the proper frequency for the CDI:
\begin{equation}
\label{freq} \omega_0\approx -\frac{k_xu_{\alpha}}{2}.
\end{equation}
The present condition for physically meaningful case
$k_xu_{\alpha}/2<0$ implies that $2s+n = 0$ and $2[s+\mu]+n = 0$.

For solving Eq. (\ref{disp11}), we examine the average value of
$B_r$ with respect to $\varphi$. Then, by taking into account the
formula:
$$\frac{1}{2\pi}\int{\rm e}^{iN\varphi} d\varphi= \delta_{N,0},
$$
and preserving only the leading terms of Eq. (\ref{disp11}), after
neglecting the contribution from the plasma components, one can
derive the dispersion relation for the instability \citep{mnras}:
\begin{equation}
\label{disp} \left(\omega + \frac{k_xu_{b}}{2}\right)^2 \approx
\frac{3\omega^2_{b}k_xu_{b}}{2\gamma_{b_0}k_{\theta}c}\left[J_0\left(\frac{k_xu_{b}}
{4\Omega}\right)J_{0}\left(\frac{k_{\theta}c}{\Omega}\right)\right]^2.
\end{equation}
By expressing the frequency as $\omega\equiv\omega_0+i{\Gamma}$ it
is easy to estimate the increment of the CDI:
\begin{equation}
\label{increm} \Gamma \approx
\left(-\frac{3}{2}\frac{\omega^2_{b}}{\gamma_{b_0}}\frac{k_xu_{b}}{k_{\theta}c}\right)
^{1/2}\left|J_0\left(\frac{k_xu_{b}}
{4\Omega}\right)J_{0}\left(\frac{k_{\theta}c}{\Omega}\right)\right|.
\end{equation}

\end{document}